\documentclass[12pt]{iopart}

\usepackage{iopams}  
\usepackage{graphicx}

\begin{document}

\title[New bounds on the neutrino magnetic moment]{New bounds on the neutrino magnetic moment 
from the plasma induced neutrino chirality flip \\in a supernova}

\author{Alexander V Kuznetsov and Nickolay V Mikheev}

\address{Division of Theoretical Physics,
Department of Physics, Yaroslavl State University, 
Sovietskaya 14, 150000 Yaroslavl, Russian Federation}

\ead{avkuzn@uniyar.ac.ru, mikheev@uniyar.ac.ru}

\begin{abstract}
The neutrino chirality-flip process under the conditions of the supernova core 
is investigated in detail with the plasma polarization effects in the photon 
propagator taken into account, in a more consistent way than in earlier 
publications. It is shown in part that the contribution of the proton 
fraction of plasma is essential. New upper bounds on the neutrino magnetic 
moment are obtained: 
$\mu_\nu < (0.5 - 1.1) \, \times 10^{-12} \, \mu_{\rm B}$ 
from the limit on the supernova core luminosity for $\nu_R$ emission, 
and $\mu_\nu < (0.4 - 0.6) \, \times 10^{-12} \, \mu_{\rm B}$
from the limit on the averaged time of the neutrino spin-flip. 
The best upper bound on the neutrino magnetic moment from SN1987A 
is improved by the factor of 3 to 7.
\end{abstract}

%Uncomment for PACS numbers title message
\pacs{13.15.+g, 95.30.Cq, 97.60.Bw}
% Keywords required only for MST, PB, PMB, PM, JOA, JOB? 
%\vspace{2pc}
%\noindent{\it Keywords}: Article preparation, IOP journals
% Uncomment for Submitted to journal title message
%\submitto{\JPA}
% Comment out if separate title page not required
\maketitle

%\date{15 June 2007}

%%%%%%%%%%%%%%%%%%%%%%%%%%%%%%%%%%%%%%%%%%%%%%%%%%%%%%%%%%%%%%%%%%%%%%
\section{Introduction}
%%%%%%%%%%%%%%%%%%%%%%%%%%%%%%%%%%%%%%%%%%%%%%%%%%%%%%%%%%%%%%%%%%%%%%

Nonvanishing neutrino magnetic moment leads to various chirality-flipping 
processes where the left-handed neutrinos produced in the stellar interior
become the right-handed ones, i.e. sterile with respect to the weak 
interaction, and this can be important e.g. for the stellar 
energy-loss. 
In the standard model extended to include the neutrino mass $m_\nu$,
the well-known result for the neutrino magnetic
moment is~\cite{Lee:1977,Fujikawa:1980}:
\begin{equation}
\mu_\nu^{(SM)} = \frac{3e\,G_{\rm F} \,m_\nu}{8\pi^2\sqrt{2}} 
= 3.20 \times 10^{-19} \left(\frac{m_\nu}{1 \,\textrm{eV}}\right)
\mu_{\rm B} \,,
\label{eq:mu_nu^SM}
\end{equation}
where $\mu_{\rm B} = e / 2 m_e$ is the Bohr magneton. 
Thus, it is unobservably small given the known limits on neutrino masses. 
On the other hand, nontrivial extensions of the standard model 
such as left-right symmetry can lead to more significant values for 
the neutrino magnetic moment. 

First attempts of exploiting the mechanism of the neutrino 
chirality flipping were connected with the solar neutrino problem, and  
two different scenarios were analysed. The first one, based on the neutrino 
magnetic moment rotation in a stellar magnetic field, was investigated 
in the papers~\cite{Cisneros:1971}--\cite{Voloshin:1986b}.
In the second scenario, a neutrino changed the chirality due to the 
electromagnetic interaction of its magnetic moment with 
plasma~\cite{Clark:1973,Radomski:1975}. 
For a more extended list of references see e.g.~\cite{Raffelt:1999}.
In all these cases the effect appeared to be small to have an essential 
impact on the solar neutrino problem, if $\mu_\nu < 10^{-10} \,\mu_{\rm B}$. 

A more significant constraints on $\mu_\nu$ are provided by other stars. 
For example, the cores of low-mass red giants are about $10^4$ times denser 
than the Sun, and nonstandard neutrino losses would have a more essential 
effect there, delaying the ignition of heluim. Thus, the limit was 
obtained~\cite{Raffelt:1990a,Raffelt:1990b}: 
\begin{eqnarray}
\mu_\nu < 0.3 \times 10^{-11} \, \mu_{\rm B} \,.
\label{eq:lim_Raff90}
\end{eqnarray}

An independent constraint on the magnetic moment of a neutrino was also 
obtained from the Early Universe~\cite{Fukugita:1987,Elmfors:1997}:
\begin{eqnarray}
\mu_\nu < 6.2 \times 10^{-11} \, \mu_{\rm B} \,,
\label{eq:lim_Elmf}
\end{eqnarray}
where spin-flip collisions would populate the sterile Dirac components 
in the era before the decoupling of the neutrinos. 
Thus, it doubles the effective number of thermally excited neutrino degrees 
of freedom and increases the expansion rate of the Universe, causing the 
overabundance of helium.

A considerable interest to the neutrino magnetic moment arised after 
the great event of SN1987A, 
in connection with the modelling of a supernova explosion, where 
gigantic neutrino fluxes define in fact the process energetics. 
It means that such  a microscopic neutrino characteristic, as the neutrino 
magnetic moment, would have a critical influence on macroscopic properties 
of these astrophysical events. Namely, the left-handed neutrinos produced 
inside the supernova core during the collapse, could convert into the 
right-handed neutrinos due to the magnetic moment interaction. 
These sterile neutrinos would escape from the core leaving no energy 
to explain the observed neutrino luminosity 
of the supernova. Thus, the upper bound on the neutrino magnetic moment 
can be established. 

This matter was investigated by many authors in different aspects
~\cite{Goldman:1988}--\cite{Goyal:1995}. 
We will mainly focus on the paper by R. Barbieri and 
R.~N. Mohapatra~\cite{Barbieri:1988} which now looks as the most 
reliable instant constraint on the neutrino magnetic moment
from SN1987A, according to~\cite{RevPartPhys:2006}. 
The authors~\cite{Barbieri:1988} considered the neutrino spin-flip via both 
$\nu_L e^- \to \nu_R e^-$
and $\nu_L p \to \nu_R p$ scattering processes in the inner core of a supernova 
immediately after the collapse. 
Imposing for the $\nu_R$ luminosity $Q_{\nu_R}$ the 
upper limit of $10^{53}$ ergs/s, the authors 
obtained the upper bound on the neutrino magnetic moment:
\begin{eqnarray}
\mu_\nu < (0.2-0.8) \times 10^{-11} \, \mu_{\rm B} \,.
\label{eq:lim_Barbi}
\end{eqnarray}

However, the essential plasma polarization effects in the photon propagator 
were not considered in~\cite{Barbieri:1988}, and the photon dispersion 
was taken in a phenomenolical way, by inserting an {\it ad hoc} thermal mass 
into the vacuum photon propagator. 
A detailed investigation of this question 
was performed in the papers by A. Ayala, J.~C. D'Olivo and 
M. Torres~\cite{Ayala:1999,Ayala:2000}, who 
used the formalism of the thermal field theory to take into account 
the influence of hot dense astrophysical plasma on the photon propagator. 
The upper bound on the neutrino magnetic moment compared with the result 
of the paper~\cite{Barbieri:1988}, was improved in~\cite{Ayala:1999,Ayala:2000} 
in the factor of 2:
\begin{eqnarray}
\mu_\nu < (0.1-0.4) \times 10^{-11} \, \mu_{\rm B} \,.
\label{eq:lim_Ayala}
\end{eqnarray}
However, looking at the intermediate analytical results by the 
authors~\cite{Ayala:1999,Ayala:2000}, one can see
that only the contribution of plasma electrons was taken 
into account there, while the proton fraction was omitted.
This is despite the fact that the electron and proton contributions to the 
neutrino spin-flip process were evaluated in~\cite{Barbieri:1988} to be 
of the same order.
Thus, the reason exists to reconsider the neutrino spin-flip processes 
in the supernova core more attentively. 

In this paper, we perform such an analysis, and
we show in part, that the proton contribution into the photon propagator 
is not less essential, than the electron contribution. 

We consider the Dirac neutrinos only, because in this case the neutrino 
magnetic moment interaction (both diagonal and non-diagonal) with a photon 
transforms the active left-handed neutrinos into the right-handed neutrinos 
which are sterile with respect to the weak interaction. 
We do not consider the Majorana neutrinos, because the produced right-handed 
antineutrino states are not sterile in this case. 

We begin in Sec.~\ref{sec:background} with calculations of the amplitude 
of the neutrino spin-flip process due to the neutrino scattering off plasma 
components. We formulate a general expression for 
the rate of creation of the right-handed neutrino with the fixed energy. 
Some details of calculations are presented 
in~\ref{app:eigenvalues} and~\ref{app:integration}. 
In Sec.~\ref{sec:luminosity} we calculate 
the supernova core luminosity for $\nu_R$ emission and we obtain 
the upper limit on the neutrino magnetic moment. 
Another possibility of imposing the upper limit on the neutrino magnetic 
moment from estimation of the averaged time of the 
left-handed neutrino washing away, i.e. of the total 
conversion of left-handed neutrinos to right-handed neutrinos  
is also considered. 
In Sec.~\ref{sec:summary} we give our conclusions.

%%%%%%%%%%%%%%%%%%%%%%%%%%%%%%%%%%%%%%%%%%%%%%%%%%%%%%%%%%%%%%%%%%%%%%
\section{Neutrino interaction with background}
\label{sec:background}
%%%%%%%%%%%%%%%%%%%%%%%%%%%%%%%%%%%%%%%%%%%%%%%%%%%%%%%%%%%%%%%%%%%%%%

%%%%%%%%%%%%%%%%%%%%%%%%%%%%%%%%%%%%%%%%%%%%%%%%%%%%%%%%%%%%%%%%%%%%%%
\subsection{The neutrino chirality flip amplitude}
\label{subsec:amplitude}
%%%%%%%%%%%%%%%%%%%%%%%%%%%%%%%%%%%%%%%%%%%%%%%%%%%%%%%%%%%%%%%%%%%%%%

The neutrino chirality flip is caused by the scattering via the 
intermediate photon (plasmon) off the plasma electromagnetic current presented 
by electrons, $\nu_L e^- \to \nu_R e^-$, protons, $\nu_L p \to \nu_R p$, etc. 
The total process Lagrangian consists of two parts, the first one is the 
interaction of a neutrino having a magnetic moment $\mu^{i j}_\nu$ 
(both diagonal and transition) with photons, 
while the second part describes the plasma interaction with photons:
\begin{equation}
{\cal L} = - \frac{\mathrm{i}}{2}  \, \sum\limits_{i,j} \mu^{i j}_\nu
\left( \bar \nu_j \sigma_{\alpha \beta} 
\nu_i \right) F^{\alpha \beta} 
- e \, J_\alpha \, A^\alpha \,, 
\label{eq:L}
\end{equation}
where $\sigma_{\alpha \beta} = (1/2)\, (\gamma_\alpha \gamma_\beta - 
\gamma_\beta \gamma_\alpha)$, 
$F^{\alpha \beta}$ is the tensor of the photon electromagnetic field, 
$e > 0$ is the elementary charge, 
$J_\alpha = 
- \left( \bar e \gamma_\alpha e \right) 
+ \left( \bar p \gamma_\alpha p \right) 
+ \dots$
is an electromagnetic current in the general 
sense, formed by different components of the medium, i.e. free 
electrons and positrons, 
protons, free ions, etc. 
Here we will consider the diagonal neutrino magnetic moment $\mu_\nu$. 
An extension to the case of the transition magnetic moment 
$\mu^{i j}_\nu$ is straightforward. 

\begin{figure}[htb]
\centering
\includegraphics[width=0.18\textwidth]{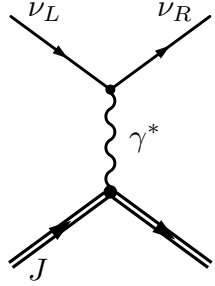}
\caption{The Feynman diagram for the neutrino spin-flip scattering via the 
intermediate plasmon $\gamma^*$ on the plasma electromagnetic current 
$J$.}
\label{fig:inter_plasmon}
\end{figure}

With the Lagrangian~(\ref{eq:L}), the process is described by the Feynman 
diagram shown in Fig.~\ref{fig:inter_plasmon}.

%%%%%%%%%%%%%%%%%%%%%%%%%%%%%%%%%%%%%%%%%%%%%%%%%%%%%%%%%%%%%%%%%%%%%%%%%
%%%%%%%%%%%%%%%%%%%%%%%%%%%%%%%%%%%%%%%%%%%%%%%%%%%%%%%%%%%%%%%%%%%%%%%%%

The technics of calculations of the neutrino spin-flip rate is rather standard.
The invariant amplitude for the process of the neutrino scattering off the 
$k$-th plasma component can be written in the form
\begin{equation}
{\cal M}^{(k)} = - 
\mathrm{i} \,e \,\mu_\nu \, j^\alpha_{(\nu)} \, 
G_{\alpha \beta}  (Q)\,J^\beta_{(k)} \,,
\label{eq:M}
\end{equation}
where $j^\alpha_{(\nu)}$ is the Fourier transform of the neutrino magnetic 
moment current, 
$$j^\alpha_{(\nu)} = \left[ \bar \nu_R (p ') \, \sigma^{\mu \alpha} \,
\nu_L (p) \right] Q_\mu \,,$$
$J^\beta_{(k)}$ is the Fourier transform of the $k$-th plasma component  
electromagnetic current, and $Q = (q_0, {\bf q})$ is the four-momentum 
transferred. The only principal point is to use 
the photon propagator $G_{\alpha \beta} (Q)$ with the plasma polarization 
effects taken into account. 
We use the straightforward way of taking account of these effects 
by summation of the 
Feynman diagrams of the forward photon scattering off plasma particles. 
Similarly to the vacuum case, this summation leads to the Dyson equation 
which provides a correct result for the photon propagator in plasma 
in the region where the photon polarization operator is real, 
in the form:

\begin{eqnarray}
G_{\alpha \beta} (Q) = 
\frac{\mathrm{i} \,\rho_{\alpha \beta} (t)}{Q^2 - \Pi_t} +
\frac{\mathrm{i} \,\rho_{\alpha \beta} (\ell)}{Q^2 - \Pi_{\ell}} \,,
\label{eq:G_alpha_beta}
\end{eqnarray}
where 
$\Pi_{t,\ell}$ are the eigenvalues of the photon polarization tensor 
$\Pi_{\alpha \beta}$ for the transverse and longitudinal plasmon, 
\begin{eqnarray}
\Pi_{\alpha \beta} = - \Pi_t \, \rho_{\alpha \beta} (t) 
- \Pi_{\ell} \, \rho_{\alpha \beta} (\ell) \,, 
\label{eq:Pi_alpha_beta}
\end{eqnarray}
and $\rho_{\alpha \beta} (t,\ell)$ are the corresponding density 
matrices 

\begin{eqnarray}
\rho_{\alpha \beta} (t) &=& - \left( g_{\alpha \beta} - \frac{Q_\alpha Q_\beta}{Q^2} 
- \frac{L_\alpha L_\beta}{L^2} \right),
\label{eq:rho_t}\\
\rho_{\alpha \beta} (\ell) &=& - \frac{L_\alpha L_\beta}{L^2} \,,
\label{eq:rho_l}\\
L_\alpha &=& Q_\alpha \, (u \, Q) - u_\alpha \, Q^2 \,,
\label{eq:vec_l} 
\end{eqnarray}
$u_\alpha$ is the four-vector of the plasma velocity. 
The density matrices $\rho_{\alpha \beta} (\lambda)$ with $\lambda = t, \ell$ 
have properties of the projection operators:
\begin{eqnarray}
\rho_{\alpha \mu} (\lambda) \, \rho^\mu_\beta (\lambda ') = 
- \delta_{\lambda \lambda '} \, \rho_{\alpha \beta} (\lambda) \,.
\label{eq:project_oper}
\end{eqnarray}

In the region where the eigenvalues $\Pi_{t,\,\ell}$
of the photon polarization tensor develop imaginary parts, they 
can be written as:
\begin{eqnarray}
\Pi_\lambda = R_\lambda + \mathrm{i} \, I_\lambda \,, 
\label{eq:Pi}
\end{eqnarray}
where $R_\lambda$ and $I_\lambda$ are the real and imaginary parts, 
containing the contributions of all components of the active medium.
For extracting the imaginary parts $I_{t,\,\ell}$, it will suffice  
to make an analytical extension $q_0 \to q_0 + \mathrm{i} \, \epsilon$
corresponding to the retarded polarization operator. 

The eigenvalues $\Pi_{t,\,\ell}$
of the photon polarization tensor are presented 
in~\ref{app:eigenvalues} both in the general form and 
in some particular cases. 

%%%%%%%%%%%%%%%%%%%%%%%%%%%%%%%%%%%%%%%%%%%%%%%%%%%%%%%%%%%%%%%%%%%%%%
\subsection{The rate of creation of the right-handed neutrino}
\label{subsec:spectrum_function}
%%%%%%%%%%%%%%%%%%%%%%%%%%%%%%%%%%%%%%%%%%%%%%%%%%%%%%%%%%%%%%%%%%%%%%

The value of physical interest is the rate of creation of the right-handed 
neutrino $\nu_R$, $\Gamma_{\nu_R} (E ')$, with the fixed energy $E '$ by 
all the left-handed neutrinos. This function can be obtained by 
integration of the amplitude~(\ref{eq:M}) squared 
over the states of the initial left-handed neutrinos and 
over the states of the initial and final plasma particles forming 
the electromagnetic current $J^\beta_{(k)}$

\begin{eqnarray}
\Gamma_{\nu_R} (E ') &=& \sum\limits_k \Gamma^{(k)}_{\nu_R} (E ') \,,
\label{eq:Gamma_def0}\\[2mm]
\Gamma^{(k)}_{\nu_R} (E ') &=& \frac{1}{16 \, (2 \pi)^5 \, E '} \, \int \, 
\sum\limits_{s, s'} \, |{\cal M}^{(k)}|^2 \, 
\delta^{(4)} (p ' + {\cal P} ' - p - {\cal P}) \,
\nonumber\\[2mm]
&\times& 
\frac{\mathrm{d}^3 {\bf P}}{\cal E} \, f_k ({\cal E}) \,
\frac{\mathrm{d}^3 {\bf P} '}{{\cal E} '} \, \left[1 \mp f_k ({\cal E} ') \right]
\frac{\mathrm{d}^3 {\bf p}}{E} \, f_\nu (E) \,.\, 
\label{eq:Gamma_def1}
\end{eqnarray}
Here, $p^\alpha = (E, {\bf p})$ and $p '^\alpha = (E ', {\bf p} ')$ 
are the four-momenta of the initial and final neutrinos, 
${\cal P}^\alpha = ({\cal E}, {\bf P})$ and
${\cal P} '^\alpha = ({\cal E} ', {\bf P} ')$ 
are the four-momenta of the initial and final plasma particles; 
$\sum_{s, s'}$ means the summation over the spins of these particles, 
the index $k = e, p, i, \dots$ corresponds to the type of the plasma 
particles (electrons, protons, free ions, etc.) 
with the distribution function $f_k ({\cal E})$, which can be 
both fermions (the upper sign in $\left[1 \mp f_k ({\cal E} ') \right]$) 
and bosons (the lower sign); 
$f_\nu (E) = \left(\mathrm{e}^{(E - \tilde \mu_\nu)/T} + 1\right)^{-1}$ 
is the Fermi--Dirac distribution function for the initial left-handed neutrinos 
in the plasma restframe, $\tilde \mu_\nu$ is the neutrino chemical potential. 

It is convenient to pass in Eq.~(\ref{eq:Gamma_def1}) from integration over 
the initial neutrino momentum ${\bf p}$ to the integration over 
the virtual plasmon momentum $p - p ' = Q = (q_0, {\bf q}), \, 
|{\bf q}| \equiv q$, using the relation:
$$\frac{\mathrm{d}^3 {\bf p}}{E} \, f_\nu (E) = \frac{2 \, \pi}{E '} \,
q \, \mathrm{d} q \, \mathrm{d} q_0 \, \theta (- Q^2) \, 
\theta (2 E ' + q_0 - q) \, f_\nu (E ' + q_0) \,.$$

Substituting the amplitude~(\ref{eq:M}) squared into 
Eq.~(\ref{eq:Gamma_def1}), one obtains
\begin{eqnarray}
\Gamma_{\nu_R} (E ') &=& \frac{\mu_\nu^2}{8 \, \pi^2 \, E '^2} \,
\int\limits_{-E '}^\infty \, \mathrm{d} q_0 \, 
\int\limits_{|q_0|}^{2 E ' + q_0} \, q \, \mathrm{d} q 
\, f_\nu (E ' + q_0) \, 
j_{(\nu)}^\alpha \, j_{(\nu)}^{\alpha ' *} \,
\nonumber\\[2mm]
&\times& 
\sum\limits_{\lambda, \lambda '} \, \frac{\rho_{\alpha \beta} (\lambda) \,
\rho_{\alpha ' \beta '} (\lambda ')}{(Q^2 - \Pi_{\lambda}) \, 
(Q^2 - \Pi_{\lambda '}^*)} \; T^{\beta \beta '} \,,
\label{eq:Gamma_def_T}
\end{eqnarray}
where the following tensor integral is introduced:
\begin{eqnarray}
T^{\alpha \beta} &=& \frac{e^2}{32 \, \pi^2} \, \sum\limits_k \, 
\sum_{s, s'} \int \, J^\alpha_{(k)} J^{\beta *}_{(k)} \; \mathrm{d}\, \Phi 
\, ,
\label{eq:T_int}\\
\mathrm{d}\, \Phi &=& 
\frac{\mathrm{d}^3 {\bf P} \, \mathrm{d}^3 {\bf P} '}{{\cal E} \, {\cal E} '}\, 
f_k ({\cal P}) \left[1 \mp f_k ({\cal P} ') \right] 
\delta^{(4)} ({\cal P} ' - {\cal P} - Q)\,.
\nonumber
\end{eqnarray}

The detailed calculation of the tensor $T^{\alpha \beta}$ is presented 
in~\ref{app:integration}. It is remarkable that the result is expressed in 
terms of the density matrices~(\ref{eq:rho_t}), (\ref{eq:rho_l}): 
\begin{eqnarray}
T^{\alpha \beta} = \left[ - I_t \, \rho^{\alpha \beta} (t) - 
I_\ell \, \rho^{\alpha \beta} (\ell) \right] \left[ 1 + f_\gamma (q_0) \right],
\label{eq:T_int_res}
\end{eqnarray}
where $I_{t, \, \ell}$ are the imaginary parts of the eigenvalues 
$\Pi_{t,\,\ell}$ of the photon polarization tensor; 
$f_\gamma (q_0)$ is the Bose--Einstein distribution function for a photon. 

Substituting~(\ref{eq:T_int_res}) into~(\ref{eq:Gamma_def_T}), using the  
orthogonality of the tensors $\rho^{\alpha \beta} (t)$ and 
$\rho^{\alpha \beta} (\ell)$, see Eq.~(\ref{eq:project_oper}), 
and taking into account the expressions for the contracts of the 
neutrino current with these tensors:
$$ j^\alpha_{(\nu)} \, j^{\beta *}_{(\nu)} \, \rho_{\alpha \beta} (t) = 
Q^4 \left[ \frac{(2 E ' + q_0)^2}{q^2} - 1 \right] , $$
$$ j^\alpha_{(\nu)} \, j^{\beta *}_{(\nu)} \, \rho_{\alpha \beta} (\ell) = 
- Q^4 \, \frac{(2 E ' + q_0)^2}{q^2} \,, $$
one finally obtains for the rate of creation of the right-handed neutrino: 
\begin{eqnarray}
&&\Gamma_{\nu_R} (E ') = \frac{\mu_\nu^2}{16\, \pi^2 \, E '^2} \; 
\int\limits_{-E '}^\infty \, \mathrm{d} q_0 \, 
\int\limits_{|q_0|}^{2 E ' + q_0} \, q^3 \, \mathrm{d} q 
\, f_\nu (E ' + q_0) \, (2 E ' + q_0)^2 \,
\nonumber\\
&&\times 
\left( 1 - \frac{q_0^2}{q^2} \right)^2 \left[ 1 + f_\gamma (q_0) \right]
\left[ 
\left( 1 - \frac{q^2}{(2 E ' + q_0)^2} \right)\, \varrho_t -  
\varrho_\ell \right] .
\label{eq:Gamma_main}
\end{eqnarray}
Here, the plasmon spectral densities are introduced:
\begin{eqnarray}
\varrho_\lambda = \frac{- 2 \, I_\lambda}{(Q^2 - R_\lambda)^2 
+ I_\lambda^2}\,,
\label{eq:varrho_t_ell}
\end{eqnarray}
which are defined by the eigenvalues~(\ref{eq:Pi})
of the photon polarization tensor~(\ref{eq:Pi_alpha_beta}).

The formula~(\ref{eq:Gamma_main}) presents our main result. 
We note that it is in agreement, to the notations, with the rate obtained 
by P. Elmfors et al.~\cite{Elmfors:1997} from the retarded self-energy 
operator of the right-handed neutrino.  
However, extracting from our general expression the electron contribution 
only, we obtain the result which is larger by the factor of 2 
than the corresponding formula in the papers by A. Ayala 
et al.~\cite{Ayala:1999,Ayala:2000}. 
It can be seen that an error was made there just in the first formula 
defining the production rate $\Gamma$ of a right-handed neutrino. 

Our formula being obtained for the process of the neutrino interaction with 
virtual photons, has in fact a more general sense, 
and can be used for neutrino-photon 
processes in any optically active medium. We only need to identify 
the photon spectral density functions $\varrho_\lambda$. 
For example, in the medium where $I_t \to 0$ 
in the space-like region $Q^2 < 0$ corresponding to the refractive 
index values $n > 1$, the spectral density function is transformed to 
$\delta$-function, and we can reproduce the result of the paper by 
W. Grimus and H. Neufeld~\cite{Grimus:1993} devoted to the study of the 
Cherenkov radiation of transversal photons by neutrinos. 

If one formally takes the limit 
$I_\ell \to 0$, 
the result obtained by S. Mohanty and S. Sahu~\cite{Mohanty:1997} can be 
reproduced, namely, the width of the Cherenkov radiation and absorption of 
longitudinal photons by neutrinos in the space-like region 
$Q^2 < 0$. However, the limit $I_\ell \to 0$ itself is irrelevant 
for $Q^2 < 0$ 
in the real astrophysical plasma conditions considered by those authors 
and leads to the strong overestimation of a result.

%%%%%%%%%%%%%%%%%%%%%%%%%%%%%%%%%%%%%%%%%%%%%%%%%%%%%%%%%%%%%%%%%%%%%%
\subsection{Contributions of plasma components into the neutrino 
scattering process}
\label{subsec:contributions}
%%%%%%%%%%%%%%%%%%%%%%%%%%%%%%%%%%%%%%%%%%%%%%%%%%%%%%%%%%%%%%%%%%%%%%

As it was mentioned above, an analysis of the neutrino chirality flip 
process has to be performed with taking account of the neutrino scattering 
off various plasma components: electrons, protons, free ions, etc.
For the first step we consider the contribution of the neutrino scattering 
off electrons into the right-handed neutrino production rate.
This means that we take into account the electron contribution only 
into the function $I_\lambda$ in the numerator of Eq.~(\ref{eq:varrho_t_ell}). 
It should be stressed however, that the functions $R_\lambda$ and $I_\lambda$ 
in the denominator of Eq.~(\ref{eq:varrho_t_ell}) contain the contributions 
of all plasma components. At this point our result for the neutrino scattering 
off electrons differs from the result 
by A. Ayala et al.~\cite{Ayala:1999,Ayala:2000}, where the electron 
contribution only was taken both in the numerator and in the denominator of the 
plasmon spectral densities.  

As the analysis shows, see~\ref{app:eigenvalues}, the electron and proton 
contributions into the imaginary parts $I_\lambda$ of the eigenvalues 
$\Pi_\lambda$ of the photon polarization tensor are of the same order of 
magnitude and have the same sign both for $\lambda = t$ and for 
$\lambda = \ell$, see Figs.~\ref{fig:ImL} and~\ref{fig:ImT}. 
This fact itself 
should lead to a decreasing of the electron contribution into the 
function $\Gamma_{\nu_R} (E ')$. On the other hand, it is seen from 
Fig.~\ref{fig:ReL}, that the electron and proton 
contributions into the real part $R_{\ell}$ of the eigenvalue 
$\Pi_{\ell}$ are of the same order of magnitude but have the opposite signs 
in the region where the imaginary part of the electron contribution into 
the numerator of Eq.~(\ref{eq:varrho_t_ell}) is relatively large. 
As a result, the contribution of the neutrino scattering 
off electrons into the right-handed neutrino production rate, obtained 
by us, appears to be close to the result by A. Ayala et al., besides the 
above-mentioned factor of 2. 

It is possible to consider similarly the contribution of 
the neutrino scattering 
off protons into the right-handed neutrino production rate. In this case, 
we take the proton contribution into the functions $I_\lambda$ 
~(\ref{eq:Itp}),~(\ref{eq:Ilp}) in the numerator 
of Eq.~(\ref{eq:varrho_t_ell}). 

\begin{figure}[htb]
\centering
\includegraphics[width=0.8\textwidth]{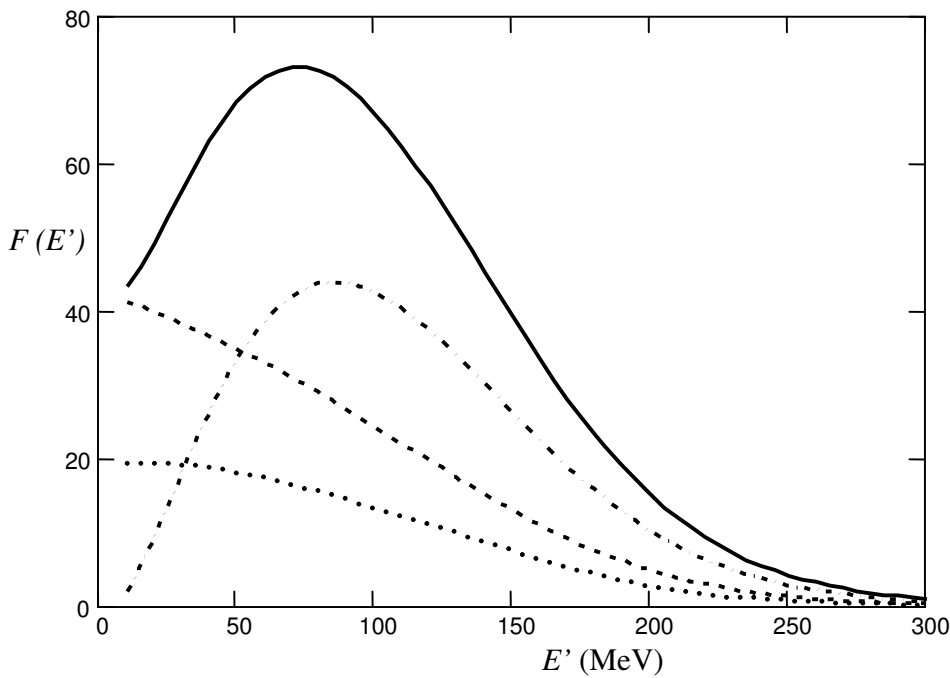}
\caption{The function $F (E ')$ defining the electron contribution 
(dashed line), the proton contribution (dash-dotted line) into the 
$\nu_R$ production rate, and the total rate (solid line)
for the plasma temperature $T =$30 MeV. 
The dotted line shows the result by A. Ayala et al.~\cite{Ayala:2000}.}
\label{fig:rate_fuction}
\end{figure}

The results of our numerical analysis of the separate contributions of 
the neutrino scattering off electrons and protons, as well as the total 
$\nu_R$ production rate in the typical conditions of the supernova core 
are presented in Fig.~\ref{fig:rate_fuction}. 

%%%%%%%%%%%%%%%%%%%%%%%%%%%%%%%%%%%%%%%%%%%%%%%%%%%%%%%%%%%%%%%%%%%%%%%%%
%%%%%%%%%%%%%%%%%%%%%%%%%%%%%%%%%%%%%%%%%%%%%%%%%%%%%%%%%%%%%%%%%%%%%%%%%

The plotted function $F (E ')$ is defined by the expression
\begin{eqnarray}
\Gamma_{\nu_R} (E ') = \frac{\mu_\nu^2\ T^3}{32\ \pi}\, F (E ')\,.
\label{eq:F(E)_def}
\end{eqnarray}
For comparison, the result by A. Ayala et al.~\cite{Ayala:2000} is also 
shown in Fig.~\ref{fig:rate_fuction}, illustrating a strong underestimation 
of the neutrino chirality flip rate made by those authors. 

We consider also the contribution of the neutrino scattering 
off free ions into the $\nu_R$ production rate. While the ions are 
believed to be absent in the supernova core, a significant fraction of them 
could be presented e.g. in the upper layers of the supernova envelope. 
It should be mentioned that longitudinal virtual plasmons give the main 
contribution into the $\nu_R$ production rate in this case. 
As is seen from Eqs.~(\ref{eq:R,Ii}), the 
function $I^{(i)}_{\ell}$ differs from zero only in the 
narrow area $\Delta x$ of the variable $x = q_0/q$, namely, 
$\Delta x\sim \sqrt{T/m_i} \ll 1$, where $m_i$ is the ion mass. 
This allows to perform calculations of the ion contribution into the 
$\nu_R$ production rate analitically, to obtain:
\begin{eqnarray}
\Delta \, \Gamma^{(i)}_{\nu_R} (E ') = \mu_\nu^2 \, \alpha \, Z_i^2 \, n_i \, 
f_\nu (E ') 
\left( \ln \, \frac{4 E '^2 + m_{\mathrm{D}}^2}{m_{\mathrm{D}}^2} 
- \frac{4 E '^2}{4 E '^2 + m_{\mathrm{D}}^2} \right),
\label{eq:Gamma_i}
\end{eqnarray}
where $\alpha$ is the fine structure constant, $e \, Z_i$ and $n_i$ are the 
charge and the density of ions, $m_{\mathrm{D}}$ has a meaning of the Debye 
screening radius inversed, $m_{\mathrm{D}}^2 = \sum_k \, R^{(k)}_{\ell} 
(q_0 = 0)$. We remind that the summation is performed over all plasma 
components. 

It is interesting to note that Eq.~(\ref{eq:Gamma_i}) obtained in the 
approximation of heavy ions, describes rather satisfactory the proton 
contribution. 

\begin{figure}[htb]
\centering
\includegraphics[width=0.8\textwidth]{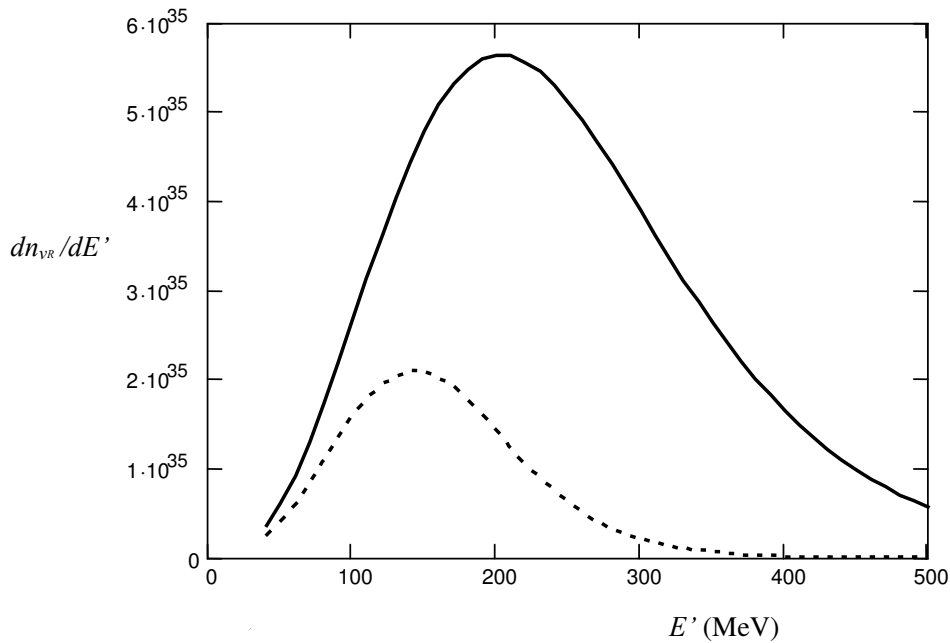}
\caption{The number of right-handed neutrinos 
(for $\mu_\nu = 10^{-12} \, \mu_{\rm B}$) 
emitted per 1 MeV of the energy spectrum per unit time from the unit volume 
for the plasma temperature $T =$ 60 MeV (solid line) and for 
$T =$ 30 MeV (dashed line).}
\label{fig:nuR_number}
\end{figure}

Given the function $\Gamma_{\nu_R} (E ')$, one can calculate the total number 
of right-handed neutrinos emitted per 1 MeV per unit time from the unit volume, 
i.e. the right-handed neutrino energy spectrum:
\begin{eqnarray}
\frac{\mathrm{d} n_{\nu_R}}{\mathrm{d} E '} = 
\frac{E '^2}{2 \, \pi^2} \, \Gamma_{\nu_R} (E ') \,.
\label{eq:dn/dE}
\end{eqnarray}
This value is presented in Fig.~\ref{fig:nuR_number} for two values 
of the plasma temperature. 

%%%%%%%%%%%%%%%%%%%%%%%%%%%%%%%%%%%%%%%%%%%%%%%%%%%%%%%%%%%%%%%%%%%%%%%%%
%%%%%%%%%%%%%%%%%%%%%%%%%%%%%%%%%%%%%%%%%%%%%%%%%%%%%%%%%%%%%%%%%%%%%%%%%

One can see from Eq.~(\ref{eq:dn/dE}), that very narrow peak of the 
function $\Gamma_{\nu_R} (E ')$ at small neutrino energy, which was analysed 
in detail in Ref.~\cite{Ayala:2000}, does not provide a huge number of 
soft right-handed neutrino production, as it was declared 
in~\cite{Ayala:2000}, because of the factor $E '^2$.

The right-handed neutrino energy spectrum~(\ref{eq:dn/dE}) can be useful for 
investigations of possible mechanisms of the energy transfer from these 
neutrinos to the outer layers of the supernova envelope. For example, 
a process is possible of the inverse conversion of a part of right-handed 
neutrinos into left-handed ones, with their subsequent absorption. Just 
these processes were proposed by A. Dar~\cite{Dar:1987} and then 
investigated in Refs.~\cite{Voloshin:1988}--\cite{Blinnikov:1988} as 
a possible mechanism for the stalled shock wave revival in the 
supernova explosion. 

%%%%%%%%%%%%%%%%%%%%%%%%%%%%%%%%%%%%%%%%%%%%%%%%%%%%%%%%%%%%%%%%%%%%%%
\section{Limits on the neutrino magnetic moment}
\label{sec:luminosity}
%%%%%%%%%%%%%%%%%%%%%%%%%%%%%%%%%%%%%%%%%%%%%%%%%%%%%%%%%%%%%%%%%%%%%%

As a possible application of the formulae obtained, we can establish the 
upper limit on the neutrino magnetic moment, by comparison of the supernova 
core luminosity computed from the $\nu_R$ energy spectrum~(\ref{eq:dn/dE}) 
with the left-handed neutrino luminosity 
$Q_{\nu_L} \sim 10^{52} - 10^{53}$ ergs/s~\cite{Raffelt:1996}, 
for a recent review see e.g.~\cite{Janka:2007}. 

The supernova core luminosity for $\nu_R$ emission can be computed as
\begin{eqnarray}
Q_{\nu_R} = V\, \int\limits_0^\infty \, \frac{\mathrm{d} n_{\nu_R}}{\mathrm{d} E '} 
\, E ' \, \mathrm{d} E ' = 
\frac{V}{2 \, \pi^2} \, \int\limits_0^\infty \, E '^3 \, \Gamma_{\nu_R} (E ') 
\, \mathrm{d} E ' \,,
\label{eq:Q_def}
\end{eqnarray}
where $V$ is the plasma volume. 

The physical conditions inside the supernova core are rather uncertain, 
they are model dependent and vary in time~\cite{Janka:2007}. 
To compare our results with the previous 
estimations~\cite{Barbieri:1988,Ayala:1999,Ayala:2000}, we use 
the same supernova core conditions as in the papers~\cite{Ayala:1999,Ayala:2000}
(plasma volume $V \sim 8 \times 10^{18} 
{\rm cm}^3$, temperature range $T = 30 - 60$ MeV, 
electron chemical potential range $\tilde \mu_e = 280 - 307$ MeV). 
These conditions could exist at the time interval before 
one second after the collapse, see~\cite{Raffelt:1996}, pp. 397-401.
We found
\begin{eqnarray}
Q_{\nu_R} = \left(\frac{\mu_\nu}{\mu_{\rm B}}\right)^2 (0.76 - 4.4) 
\times 10^{77} \; \mbox{ergs/s}\,.
\label{eq:Q_res}
\end{eqnarray}
This value should be compared with the corresponding formula (48) of 
the paper~\cite{Ayala:2000} to see that our result for the luminosity 
is greater by the factor of 10. This discrepancy can be explained by the 
following reasons: {\it i)} the factor of 2 was lost in the electron 
contribution 
in the papers~\cite{Ayala:1999,Ayala:2000}; {\it ii)} the proton contribution 
was omitted there. The neutrino scattering off protons appears to give 
even more essential contribution into the luminosity because of shift 
of the rate $\Gamma_{\nu_R} (E ')$ maximum into the region of larger 
energies, see Fig.~\ref{fig:rate_fuction}. 

Assuming that the right-handed neutrino luminosity is less than the 
left-handed neutrino lumosity at the time $\sim$ 0.1 sec after the 
collapse, $Q_{\nu_R} < 10^{53}$ ergs/s, 
we obtain from 
Eq.~(\ref{eq:Q_res}) the upper limit on the neutrino magnetic moment 
\begin{eqnarray}
\mu_\nu < (0.5 - 1.1) \, \times 10^{-12} \, \mu_{\rm B}\,.
\label{eq:mu_fr_Q}
\end{eqnarray}

An additional method can be used to put a bound on the neutrino magnetic moment.
A number of right-handed neutrinos emitted per unit time from the unit volume 
obtained by intergation of Eq.~(\ref{eq:dn/dE}) is:
\begin{eqnarray}
n_{\nu_R} = \frac{1}{2 \, \pi^2} \, \int\limits_0^\infty \, E '^2 \, 
\Gamma_{\nu_R} (E ') \, \mathrm{d} E ' \,,
\label{eq:n_def}
\end{eqnarray}
Dividing $n_{\nu_R}$ to the initial left-handed neutrino 
number density $n_{\nu_L}$, one can estimate the averaged time of the 
left-handed neutrino washing away, i.e. the deleptonization time.  
For the temperature range $T = 30 - 60$ MeV, and 
for the electron and neutrino chemical potentials $\tilde \mu_e \sim 300$ MeV, 
$\tilde \mu_\nu \sim 160$ MeV, we obtain 
\begin{eqnarray}
\tau \simeq \left(\frac{10^{-12}\,\mu_{\rm B}}{\mu_\nu}\right)^2 (0.14 - 0.36) 
 \; \mbox{sec}\,.
\label{eq:tau}
\end{eqnarray}

In order to avoid a quick deleptonization of the supernova core, 
this averaged time of the neutrino spin-flip 
should exceed a second. 
Taking the limit $\tau >$ 1 sec, we obtain the bound 
on the neutrino magnetic moment:
\begin{eqnarray}
\mu_\nu < (0.4 - 0.6) \, \times 10^{-12} \, \mu_{\rm B}\,.
\label{eq:mu_fr_num}
\end{eqnarray}

By this means, we improve the best upper bound on the neutrino magnetic 
moment from SN1987A obtained by A. Ayala et al.~\cite{Ayala:1999}
by the factor of 3 to 7.

%%%%%%%%%%%%%%%%%%%%%%%%%%%%%%%%%%%%%%%%%%%%%%%%%%%%%%%%%%%%%%%%%%%%%%
\section{Summary}                                  \label{sec:summary}
%%%%%%%%%%%%%%%%%%%%%%%%%%%%%%%%%%%%%%%%%%%%%%%%%%%%%%%%%%%%%%%%%%%%%%

We have investigated in detail the neutrino chirality-flip process under the 
conditions of the supernova core. The plasma polarization effects caused both 
by electrons and protons were taken into account in the photon propagator. 
The rate $\Gamma_{\nu_R} (E ')$ of creation of the right-handed 
neutrino with the fixed energy $E '$, the energy 
spectrum, and the luminosity have been calculated. 

From the limit on the supernova core luminosity for $\nu_R$ emission, 
we have obtained the upper bound on the neutrino magnetic moment 
$
\mu_\nu < (0.5 - 1.1) \, \times 10^{-12} \, \mu_{\rm B}\,.
$ 
From the limit on the averaged time of the neutrino spin-flip,
we have obtained the upper bound
$
\mu_\nu < (0.4 - 0.6) \, \times 10^{-12} \, \mu_{\rm B}\,.
$ 
Thus, we have improved the best upper bound on the neutrino 
magnetic moment from SN1987A by the factor of 3 to 7.

%%%%%%%%%%%%%%%%%%%%%%%%%%%%%%%%%%%%%%%%%%%%%%%%%%%%%%%%%%%%%%%%%%%%%%
\section*{ACKNOWLEDGMENTS}
%%%%%%%%%%%%%%%%%%%%%%%%%%%%%%%%%%%%%%%%%%%%%%%%%%%%%%%%%%%%%%%%%%%%%%

We thank V. A. Rubakov for useful discussion. 

The work was supported in part 
by the Council on Grants by the President of Russian Federation 
for the Support of Young Russian Scientists and Leading Scientific Schools of 
Russian Federation under the Grant No. NSh-6376.2006.2 and 
by the Russian Foundation for Basic Research under the Grant No. 07-02-00285-a. 

\appendix

%%%%%%%%%%%%%%%%%%%%%%%%%%%%%%%%%%%%%%%%%%%%%%%%%%%%%%%%%%%%%%%%%%%%%%
\section{Eigenvalues of the photon polarization tensor}
\label{app:eigenvalues}
%%%%%%%%%%%%%%%%%%%%%%%%%%%%%%%%%%%%%%%%%%%%%%%%%%%%%%%%%%%%%%%%%%%%%%

The 
expressions for the contributions of a charged fermion into the 
polarization functions $\Pi_{t,\,\ell}$ in the hard thermal loop 
approximation 
can be found e.g. in~\cite{Braaten:1993} and have the form
\begin{eqnarray}
\Pi_t &=& \frac{4 \alpha}{\pi} \, 
\int\limits_0^\infty \, \frac{\mathrm{d} {\cal P} \, {\cal P}^2}{\cal E} \, 
\left[f_F ({\cal E}) + {\bar f}_F ({\cal E}) \right]
 \left( \frac{q_0^2}{q^2} - \frac{q_0^2 - q^2}{q^2} \,
\frac{q_0}{2 v q} \ln \frac{q_0 + v q}{q_0 - v q} \right) , 
\label{eq:Pi_t}\\
\Pi_\ell &=& \frac{4 \alpha}{\pi} \, \frac{q_0^2 - q^2}{q^2} \,
\int\limits_0^\infty \, \frac{\mathrm{d} {\cal P} \, {\cal P}^2}{\cal E} \, 
\left[f_F ({\cal E}) + {\bar f}_F ({\cal E}) \right]
\nonumber\\
&\times& \left( \frac{q_0}{v q} \ln \frac{q_0 + v q}{q_0 - v q} 
- \frac{q_0^2 - q^2}{q_0^2 - v^2 q^2} - 1 \right) ,
\label{eq:Pi_el}
\end{eqnarray}
where $v = {\cal P}/{\cal E}$, and the Fermi--Dirac distribution functions for the 
fermions and anti-fermions are
\begin{eqnarray}
f_F ({\cal E}) = \frac{1}{\mathrm{e}^{({\cal E} - \tilde \mu)/T} + 1} \,, \qquad
{\bar f}_F ({\cal E}) = \frac{1}{\mathrm{e}^{({\cal E} + \tilde \mu)/T} + 1} \,, 
\label{eq:f's}
\end{eqnarray}
$\tilde \mu$ is the fermion chemical potential.

For the supernova core conditions, the main contribution comes from the plasma 
electrons and protons: 
\begin{eqnarray}
R_{t,\,\ell} \simeq R_{t,\,\ell}^{(e)} + R_{t,\,\ell}^{(p)}\,, \qquad
I_{t,\,\ell} \simeq I_{t,\,\ell}^{(e)} + I_{t,\,\ell}^{(p)}\,.
\label{eq:R,I}
\end{eqnarray}

In these conditions, there is a good approximation to consider the electron 
fraction as the relativistic plasma ($\tilde \mu_e, T \gg m_e$). 

The real and imaginary parts~(\ref{eq:R,I}) of the electron contributions into 
the photon polarization functions take the following form: 
\begin{eqnarray}
R_{t}^{(e)} &=& m_\gamma^2 \left( x^2 + \frac{x \left(1-x^2\right)}{2} \,
\ln \left|\frac{1+x}{1-x}\right|\right),
\label{eq:Rte}\\
I_{t}^{(e)} &=& - \frac{\pi}{2} \,m_\gamma^2 \,x \left(1-x^2\right) ,
\label{eq:Ite}\\
R_{\ell}^{(e)} &=& 2 \,m_\gamma^2  \left(1-x^2\right) \left( 1 - \frac{x}{2} \,
\ln \left|\frac{1+x}{1-x}\right|\right),
\label{eq:RLe}\\
I_{\ell}^{(e)} &=& \pi \,m_\gamma^2 \,x \left(1-x^2\right) ,
\label{eq:ILe}
\end{eqnarray}
where $x = q_0/q, \, |x|<1$, $m_\gamma$ is the so-called photon 
thermal mass, 
\begin{equation}
m_\gamma^2 = \frac{2 \, \alpha}{\pi} \left({\tilde \mu_e}^2 + 
\frac{\pi^2 T^2}{3} \right).
\label{eq:m_gamma}
\end{equation}

For the proton contributions, the situation appears to be more complicated. 
For the real and imaginary parts of the proton contribution into the 
polarization functions~(\ref{eq:Pi_t}), (\ref{eq:Pi_el}), for the conditions 
$\tilde \mu_p \gg T$, where $\tilde \mu_p$ is the proton chemical potential, 
one obtains:
\begin{eqnarray}
R_{t}^{(p)} &=& \frac{4 \alpha}{\pi} \, 
\int\limits_0^\infty \, \frac{\mathrm{d} {\cal P} \, {\cal P}^2}
{{\cal E} \left(\mathrm{e}^{({\cal E} - \tilde \mu_p)/T} + 1\right)} 
 \left( x^2 + \frac{x \left(1 - x^2\right)}{2 v} \, 
\ln \left|\frac{x + v}{x - v} \right| \right) ,
\label{eq:Rtp}\\
I_{t}^{(p)} &=& - 2 \alpha \, x \left(1 - x^2\right)  
\int\limits_{{\cal P}_{\rm{min}}}^\infty \frac{\mathrm{d} {\cal P} \, {\cal P}}
{\mathrm{e}^{({\cal E} - \tilde \mu_p)/T} + 1} \,, \qquad
{\cal P}_{\rm{min}} = \frac{m_p |x|}{\sqrt{1 - x^2}} \,,
\label{eq:Itp}\\
R_{\ell}^{(p)} &=& \frac{4 \alpha}{\pi} \left(1 - x^2\right) 
\int\limits_0^\infty \, \frac{\mathrm{d} {\cal P} \, {\cal P}^2}
{{\cal E} \left(\mathrm{e}^{({\cal E} - \tilde \mu_p)/T} + 1\right)} 
\nonumber\\
&\times& \left( 1 + \frac{1 - x^2}{v^2 - x^2} - \frac{x}{v} \, 
\ln \left|\frac{x + v}{x - v} \right| \right) ,
\label{eq:Rlp}\\
I_{\ell}^{(p)} &=& - 2 \, I_{t}^{(p)} + 2 \alpha \, m_p^2 \, x 
\left[ \exp \left( \frac{m_p}{T \sqrt{1 - x^2}} 
- \frac{\tilde \mu_p}{T} \right) + 1 \right]^{-1} ,
\label{eq:Ilp}
\end{eqnarray}
where $m_p$ is the effective proton mass 
in plasma~\cite{Horowitz:1987}
(in numerical calculations we take $m_p \simeq$ 700 MeV, corresponding 
to the nuclear density 3 $\times$ 10$^{14}$ g/cm$^3$).

The proton chemical potential $\tilde \mu_p$ is defined from 
the equation
\begin{eqnarray}
N_p \simeq N_e \simeq \frac{\tilde \mu_e^3}{3 \, \pi^2} = \frac{1}{\pi^2} \, 
\int\limits_0^\infty \, \frac{\mathrm{d} {\cal P} \, {\cal P}^2}
{\mathrm{e}^{({\cal E} - \tilde \mu_p)/T} + 1} \,.
\label{eq:mu_p_eq}
\end{eqnarray}
As the analysis of Eq.~(\ref{eq:mu_p_eq}) shows, the difference 
$\tilde \mu_p - m_p$ appears to be of the positive sign at the 
temperatures $T \simeq 30 - 60$ MeV, and of the same order of magnitude, 
as the temperature. Thus, in the supernova core conditions both the 
approximations of the degenerate Fermi gas and of the classical Boltzmann gas 
%appear to be hardly applicable for protons. 
should be, in general, hardly applicable for protons. 
However, we have verified by direct calculation that the observables
computed in Sec.~\ref{sec:luminosity} such as the 
luminosity~(\ref{eq:Q_def}), appear to be rather stable with respect 
to the choice of the approximation for the proton distribution function.  

\begin{figure}[htb]
\centering
\includegraphics[width=0.8\textwidth]{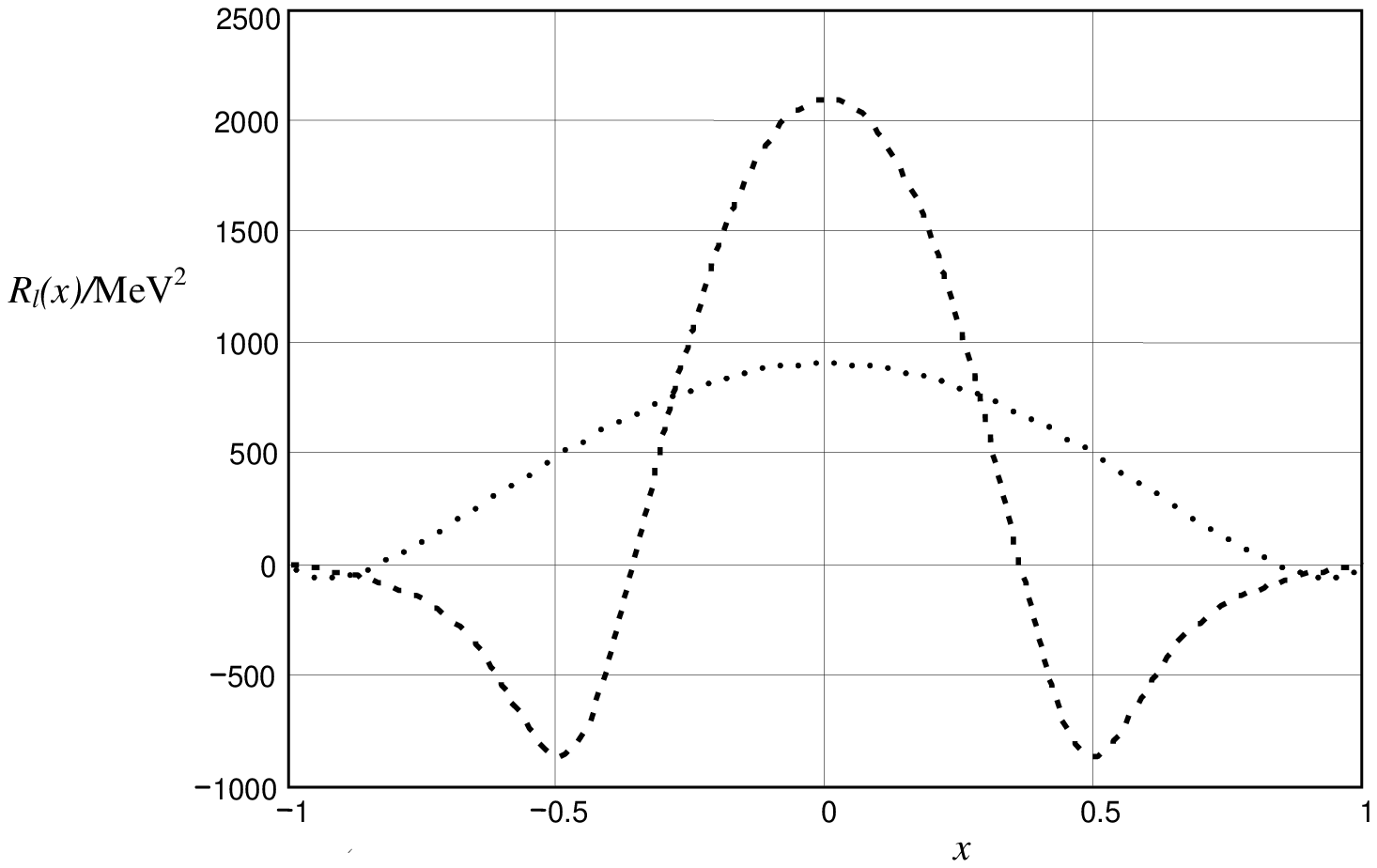}
\caption{Electron contribution (dotted line) and proton contribution 
(dashed line) at $T = 30$ MeV to the real part of $\Pi_{\ell}$.}
\label{fig:ReL}
\end{figure}

\begin{figure}[htb]
\centering
\includegraphics[width=0.8\textwidth]{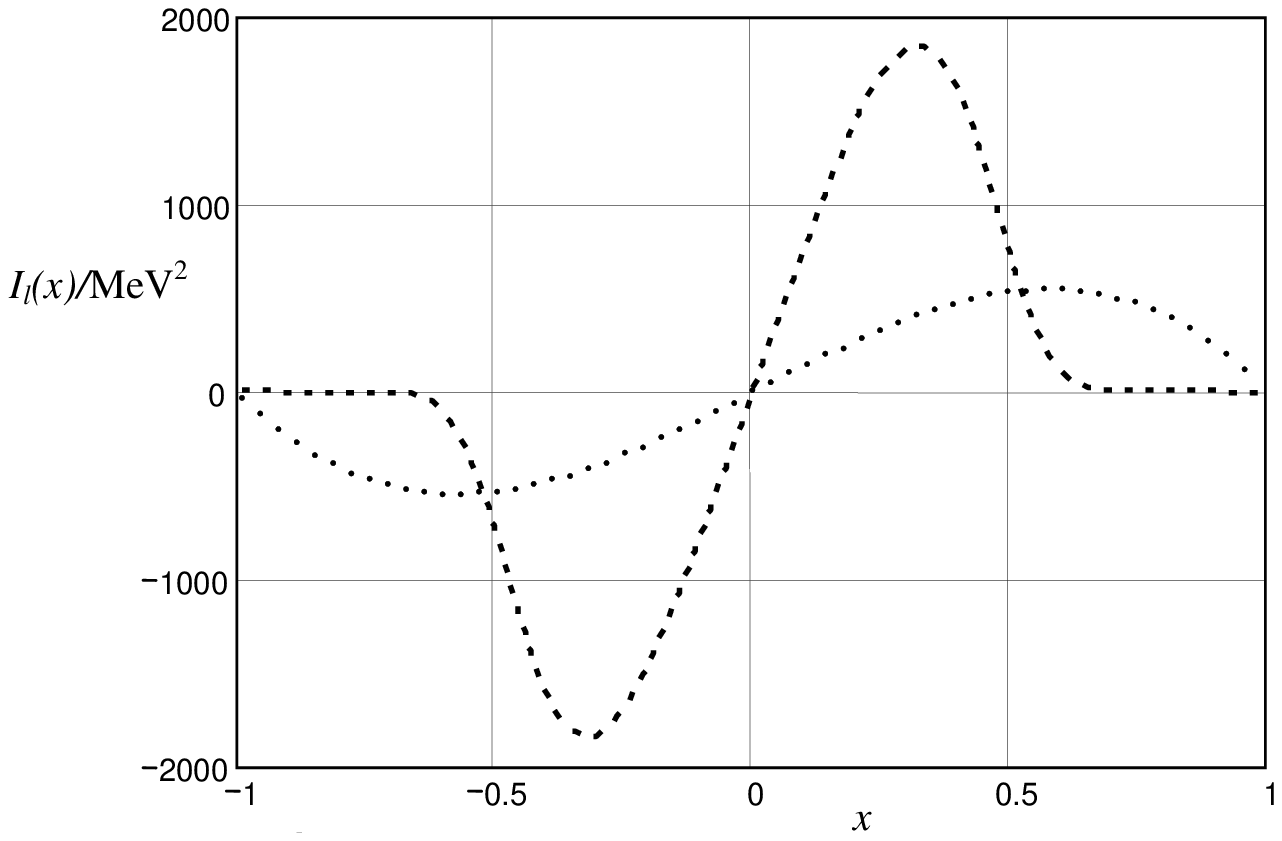}
\caption{Electron contribution (dotted line) and proton contribution 
(dashed line) at $T = 30$ MeV to the imaginary part of $\Pi_{\ell}$.}
\label{fig:ImL}
\end{figure}

\begin{figure}[htb]
\centering
\includegraphics[width=0.8\textwidth]{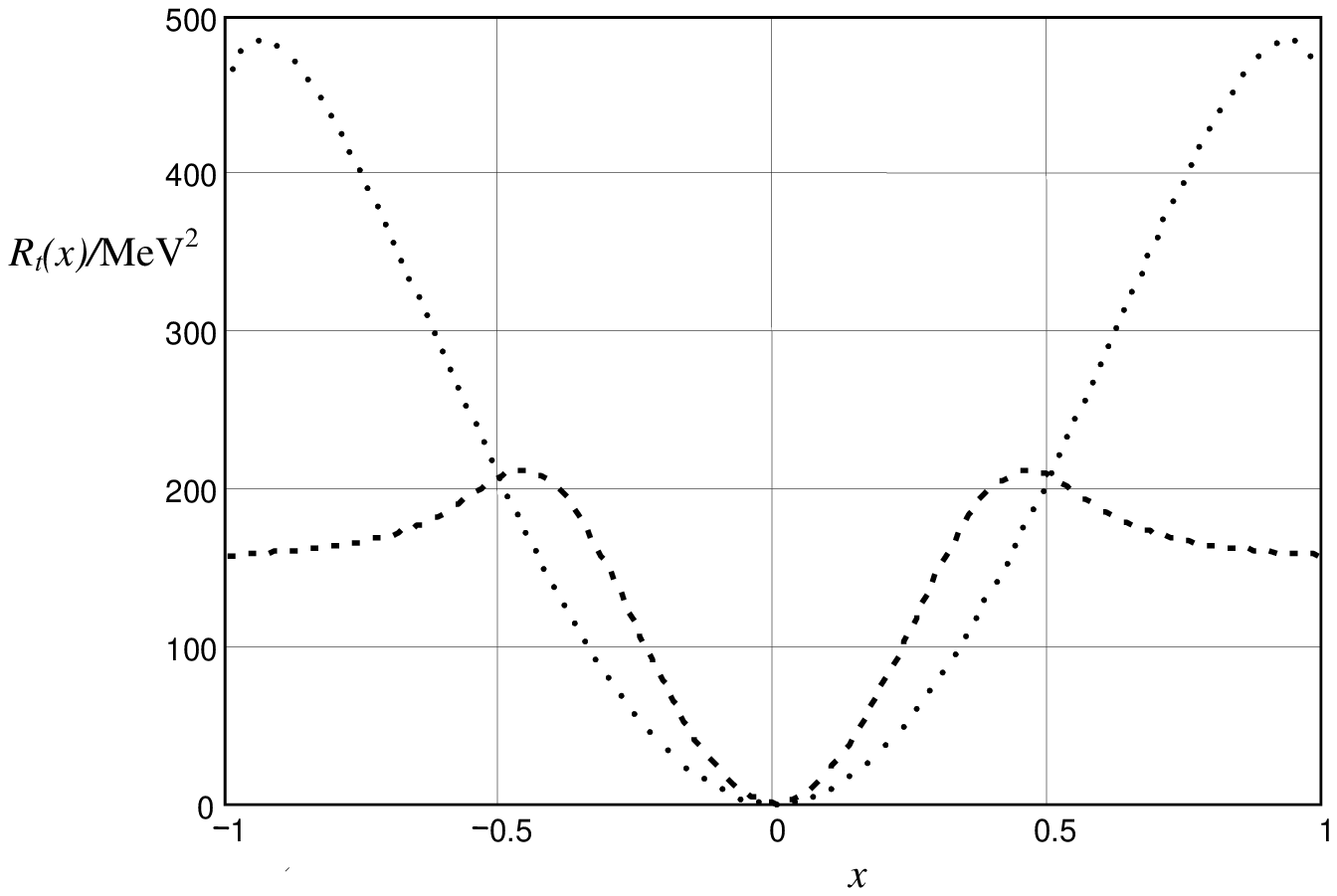}
\caption{Electron contribution (dotted line) and proton contribution 
(dashed line) at $T = 30$ MeV to the real part of $\Pi_t$.}
\label{fig:ReT}
\end{figure}

\begin{figure}[htb]
\centering
\includegraphics[width=0.8\textwidth]{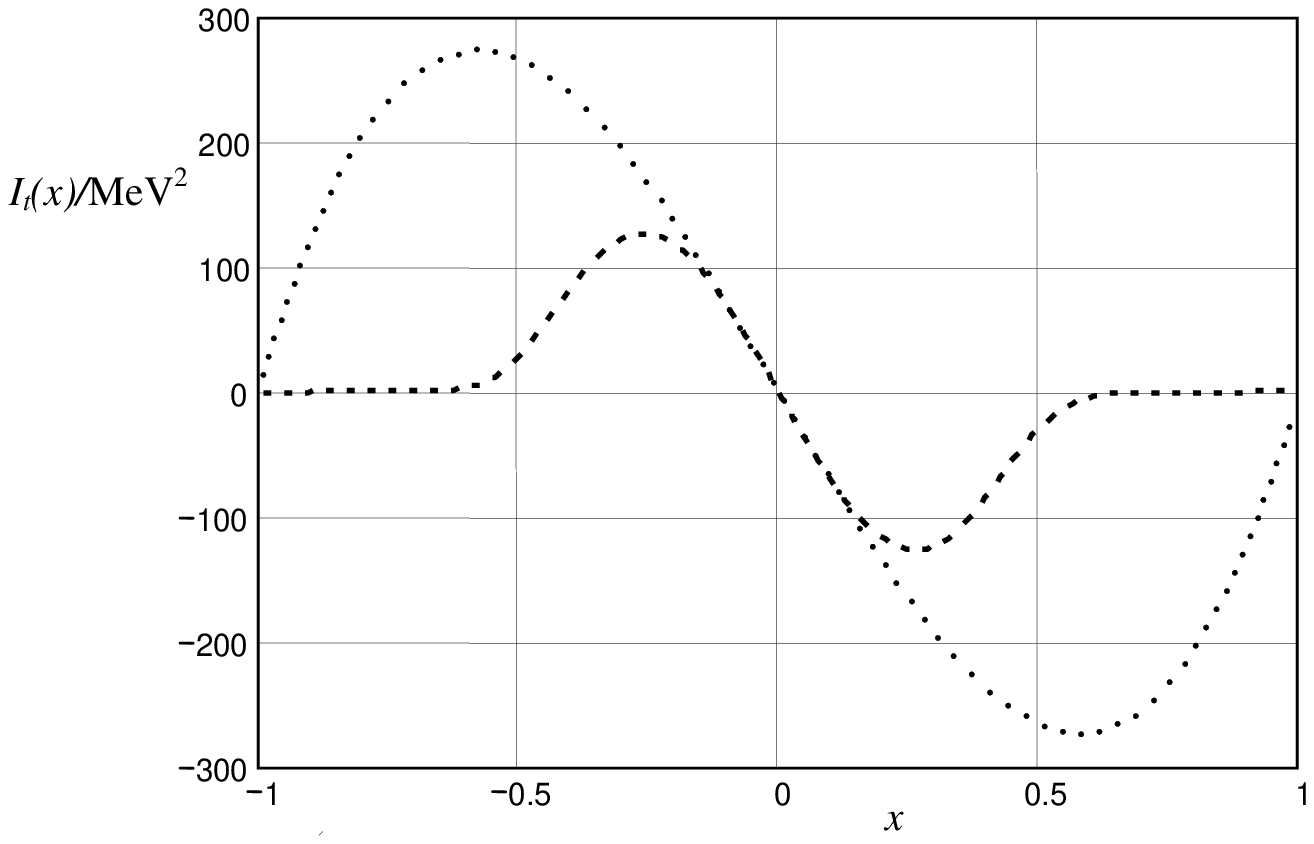} 
\caption{Electron contribution (dotted line) and proton contribution 
(dashed line) at $T = 30$ MeV to the imaginary part of $\Pi_t$.}
\label{fig:ImT}
\end{figure}

In the Figs.~\ref{fig:ReL}, \ref{fig:ImL}, \ref{fig:ReT}, 
and~\ref{fig:ImT} we present for the sake of illustration  
the electron and proton contributions 
into the eigenvalues $\Pi_{\ell, t}$ for the 
longitudinal and transverse plasmon. 
The importance of taking into account the proton contribution 
is evident. 

%%%%%%%%%%%%%%%%%%%%%%%%%%%%%%%%%%%%%%%%%%%%%%%%%%%%%%%%%%%%%%%%%%%%%%%%%
%%%%%%%%%%%%%%%%%%%%%%%%%%%%%%%%%%%%%%%%%%%%%%%%%%%%%%%%%%%%%%%%%%%%%%%%%

%%%%%%%%%%%%%%%%%%%%%%%%%%%%%%%%%%%%%%%%%%%%%%%%%%%%%%%%%%%%%%%%%%%%%%%%%
%%%%%%%%%%%%%%%%%%%%%%%%%%%%%%%%%%%%%%%%%%%%%%%%%%%%%%%%%%%%%%%%%%%%%%%%%

%%%%%%%%%%%%%%%%%%%%%%%%%%%%%%%%%%%%%%%%%%%%%%%%%%%%%%%%%%%%%%%%%%%%%%%%%
%%%%%%%%%%%%%%%%%%%%%%%%%%%%%%%%%%%%%%%%%%%%%%%%%%%%%%%%%%%%%%%%%%%%%%%%%

%%%%%%%%%%%%%%%%%%%%%%%%%%%%%%%%%%%%%%%%%%%%%%%%%%%%%%%%%%%%%%%%%%%%%%%%%
%%%%%%%%%%%%%%%%%%%%%%%%%%%%%%%%%%%%%%%%%%%%%%%%%%%%%%%%%%%%%%%%%%%%%%%%%

Together with electrons and protons, in general, a small fraction $Y_i$ 
of the free 
ions could also present in plasma. This fraction can be considered with 
a good accuracy as the classical Boltzmann gas. 
The real and imaginary parts of the corresponding polarization functions 
have the form:
\begin{eqnarray}
R_{\ell}^{(i)} &=& 4 \, \pi \, \alpha \, \frac{Z_i^2 \, n_i}{T}\, 
\left[1 -  \phi \left(\frac{x}{x_0}\right)\right],
\nonumber\\
I_{\ell}^{(i)} &=& 8 \, \pi^{3/2} \, \alpha \, Z_i^2 \, n_i \, \frac{1}{x_0 \, q}\, 
\sinh \frac{q_0}{2\, T} \, \exp \left( \frac{q^2}{8\, m_i\, T} \right) 
\exp \left( - \frac{x^2}{x_0^2} \right), 
\label{eq:R,Ii}
\end{eqnarray}
where $x_0 = \sqrt{2\, T/m_i}$, and the function is 
introduced: 
\begin{eqnarray}
\phi (y) = \frac{2}{\sqrt{\pi}} \, |y|^3 
\int\limits_0^\infty  u \, \ln \left|\frac{1+u}{1-u}\right| \, 
\mathrm{e}^{-y^2 u^2} \, \mathrm{d} u \,.
\label{eq:phi}
\end{eqnarray}

As is seen from Eq.~(\ref{eq:R,Ii}), the 
function $I^{(i)}_{\ell}$ differs from zero only in the 
narrow area of the variable $x = q_0/q$, namely, 
$ x \lesssim x_0 \sim \sqrt{T/m_i} \ll 1$. 

The functions $R_t^{(i)}$ and $I_t^{(i)}$ for the transversal plasmon 
are of the order 
$\alpha \, Z_i^2 \, n_i/m_i$ and thus are suppressed by the large 
mass of the ion in the denominator. Thus, the contribution of the neutrino 
scattering off free ions via the longitudinal plasmon ($\lambda = \ell$) 
is only essential. 

The ion contribution~(\ref{eq:R,Ii}) comes with the factor $Z_i^2 \, Y_i$, 
and it is negligibly small in the supernova core 
conditions, because of the smallness of $Y_i$. 
However, it could be essential in the upper layers of the supernova envelope, 
which are believed to be rich in iron. 

%%%%%%%%%%%%%%%%%%%%%%%%%%%%%%%%%%%%%%%%%%%%%%%%%%%%%%%%%%%%%%%%%%%%%%
\section{Integration over the initial and final plasma particles}
\label{app:integration}
%%%%%%%%%%%%%%%%%%%%%%%%%%%%%%%%%%%%%%%%%%%%%%%%%%%%%%%%%%%%%%%%%%%%%%

Here we present the detailed calculation of the tensor 
integral~(\ref{eq:T_int}):
\begin{eqnarray}
T^{\alpha \beta} &=& \frac{e^2}{32 \, \pi^2} \, \sum_k 
\sum_{s, s'} \int \, J^\alpha_{(k)} J^{\beta *}_{(k)} \; \mathrm{d}\, \Phi 
\,,
\label{eq:T_int'}\\
\mathrm{d}\, \Phi &=& 
\frac{\mathrm{d}^3 {\bf P} \, \mathrm{d}^3 {\bf P} '}{{\cal E} \, {\cal E} '}\, 
f_k ({\cal P}) \left[1 \mp f_k ({\cal P} ') \right] 
\delta^{(4)} ({\cal P} ' - {\cal P} - Q)\,.
\nonumber
\end{eqnarray}
To use the covariant properties of the tensor $T^{\alpha \beta}$, one 
should write the distribution functions $f_k ({\cal P})$ in the arbitrary frame
\begin{equation}
f_k ({\cal P}) = \left[\exp \frac{({\cal P} u) - \tilde{\mu}}{T} \pm 1 \right]^{-1},
\label{eq:f_arb}
\end{equation}
where $u_\alpha$ is the four-vector of the plasma velocity. This vector and 
the four-vector $Q_\alpha$ are the building bricks for constructing
the tensor $T^{\alpha \beta}$. This tensor is symmetric because the 
electromagnetic current $J^\alpha_{(k)}$ is real. The tensor is also orthogonal 
to the four-vector $Q_\alpha$ 
because of the electromagnetic current conservation. There exist only two 
independent structures having these properties, which are the density 
matrices~(\ref{eq:rho_t}) and~(\ref{eq:rho_l}), and thus one can write: 
\begin{equation}
T^{\alpha \beta} = {\cal A}^{(t)} \, \rho_{\alpha \beta} (t) 
+ {\cal A}^{(\ell)} \, \rho_{\alpha \beta} (\ell)\,.
\label{eq:T}
\end{equation}

Because of orthogonality of the tensors $\rho_{\alpha \beta} (t)$ and 
$\rho_{\alpha \beta} (\ell)$, see Eq.~(\ref{eq:project_oper}), 
one obtains 
\begin{eqnarray}
{\cal A}^{(t)} &=& \frac{1}{2} \, T^{\alpha \beta} \, \rho_{\alpha \beta} (t) = 
\frac{e^2}{64 \, \pi^2}  \, \rho_{\alpha \beta} (t)\, 
\sum_k \sum_{s, s'} \int \, J^\alpha_{(k)} J^{\beta *}_{(k)} 
\; \mathrm{d}\, \Phi \,,
\label{eq:A_t}\\
{\cal A}^{(\ell)} &=& T^{\alpha \beta} \, \rho_{\alpha \beta} (\ell) = 
\frac{e^2}{32 \, \pi^2} \, \rho_{\alpha \beta} (\ell) \, 
\sum_k \sum_{s, s'} \int \, J^\alpha_{(k)} J^{\beta *}_{(k)}  
\; \mathrm{d}\, \Phi \,.
\label{eq:A_l}
\end{eqnarray}

As we show below, just these integrals~(\ref{eq:A_t}) and~(\ref{eq:A_l}) 
define the widths of absorption (at $q_0 > 0$) and creation (at $q_0 < 0$) 
of a plasmon by the plasma particles. Really, let us consider for 
definiteness the width of absorption of the transversal plasmon by 
plasma particles forming the electromagnetic current $J^\beta_{(k)}$. 
The amplitude of the process has the form
\begin{equation}
{\cal M}^{(k)} (t) = - e \, \varepsilon_\alpha (t)\,J^\alpha_{(k)} \,.
\label{eq:M_abs}
\end{equation}
where $\varepsilon_\alpha (t)$ is the unit polarization four-vector. 
Performing standard calculations, one obtains for the width of the plasmon 
absorption by all the components of plasma:
\begin{equation}
\Gamma^{abs}_{(t)} = \frac{1}{32 \, \pi^2 \, q_0} \, 
\frac{1}{2} \sum_\tau \, \sum_k \, 
\sum\limits_{s, s'} \, \int \, |{\cal M}^{(k)} (t)|^2 \; \mathrm{d}\, \Phi \,,
\label{eq:Gamma_abs1}
\end{equation}
where the summation is made both over the $k$th types of the plasma particles 
and over the polarizations of all particles participating in the process, 
$\tau$ for a plasmon and $s, s'$ for plasma particles. 

Substituting the amplitude~(\ref{eq:M_abs}) into~(\ref{eq:Gamma_abs1}),
\begin{equation}
\Gamma^{abs}_{(t)} = \frac{e^2}{64 \, \pi^2 \, q_0}  \, 
\rho_{\alpha \beta} (t)\, 
\sum_k \sum_{s, s'} \int \, J^\alpha_{(k)} J^{\beta *}_{(k)} 
\; \mathrm{d}\, \Phi \,,
\label{eq:Gamma_abs2}
\end{equation}
where $\rho_{\alpha \beta} (t) = \sum\limits_{\tau = 1}^2 \, 
\varepsilon_{\alpha}^\tau (t) \, \varepsilon_\beta^\tau (t)$, 
and comparing it with Eq.~(\ref{eq:A_t}), one can find the value 
\begin{equation}
{\cal A}^{(t)} = q_0 \, \Gamma^{abs}_{(t)} \,.
\label{eq:A_t2}
\end{equation}

Using the known relation~\cite{Weldon:1983} between the width of absorption of 
the transversal plasmon and the imaginary part $I_t$ of the eigenvalue $\Pi_t$ 
of the photon polarization tensor $\Pi_{\alpha \beta}$, 
\begin{equation}
I_t (q_0) = - q_0 \left( 1 -  \mathrm{e}^{- q_0/T} \right) 
\Gamma^{abs}_{(t)} \,, 
\label{eq:I_t_gamma}
\end{equation}
we express the value ${\cal A}^{(t)}$ in terms of $I_t$: 
\begin{equation}
{\cal A}^{(t)} = - \frac{I_t}{1 -  \mathrm{e}^{- q_0/T}} = 
- I_t \left[ 1 + f_\gamma (q_0) \right],
\label{eq:A_t3}
\end{equation}
where $f_\gamma (q_0) = \left(\mathrm{e}^{q_0/T} - 1 \right)^{-1}$ 
is the Bose--Einstein distribution function for a photon. 
This relation obtained in the case $q_0 > 0$ is also correct for the case 
$q_0 < 0$, which corresponds to the transversal plasmon creation with 
the energy $\omega = - q_0 > 0$. The connection should be used here 
between the imaginary part $I_t$ and the width of creation of the transversal 
plasmon: 
\begin{equation}
I_t (\omega) = - \omega \left( \mathrm{e}^{\omega/T} - 1 \right) 
\Gamma^{cr}_{(t)} \,. 
\label{eq:I_t_gamma2}
\end{equation}
It is essential also that the function $I_t$ is odd:
\begin{equation}
I_t (- q_0) = - I_t (q_0) \,,  
\label{eq:I_t_odd}
\end{equation}
and this is the feature of the retarded polarization operator.

Performing the similar calculations, one can see that the 
relation~(\ref{eq:A_t3}) is valid for the longitudinal plasmon also. 
It is necessary to remember that $\rho_{\alpha \beta} (\ell) = 
- \varepsilon_{\alpha} (\ell) \, \varepsilon_\beta (\ell)$, and 
\begin{equation}
I_\ell (q_0) = q_0 \left( 1 -  \mathrm{e}^{- q_0/T} \right) 
\Gamma^{abs}_{(\ell)} \,. 
\label{eq:I_l_gamma}
\end{equation}

Finally, we obtain the tensor $T^{\alpha \beta}$ in the form: 
\begin{eqnarray}
T^{\alpha \beta} = \left[ - I_t \, \rho^{\alpha \beta} (t) - 
I_\ell \, \rho^{\alpha \beta} (\ell) \right] \left[ 1 + f_\gamma (q_0) 
\right]. 
\label{eq:T_int_res_app}
\end{eqnarray}
%

%%%%%%%%%%%%%%%%%%%%%%%%%%%%%%%%%%%%%%%%%%%%%%%%%%%%%%%%%%%%%%%%%%%%%%
%% References %%%%%%%%%%%%%%%%%%%%%%%%%%%%%%%%%%%%%%%%%%%%%%%%%%%%%%%%
%%%%%%%%%%%%%%%%%%%%%%%%%%%%%%%%%%%%%%%%%%%%%%%%%%%%%%%%%%%%%%%%%%%%%%

%%%%%%%%%%%%%%%%%%%%%%%%%%%%%%%%%%%%%%%%%%%%%%%%%%%%%%%%%%%%%%%%%%%%%%%

\end{document}